\documentclass[aps,prl,showpacs,twocolumn]{revtex4}
\usepackage{graphicx}
\usepackage{bbm}

\def\tr{\,{\rm tr}\,}

\def\ave#1{\langle #1 \rangle}
\def\ii{{\rm i}}

\def\tit#1{{\em #1},}
\def\etal#1{#1}

\begin{document}

\title{Exact solution for a diffusive nonequilibrium steady state of an open quantum chain}

\author{Marko \v Znidari\v c}
\address{Department of Physics, Faculty of Mathematics and Physics,
  University of Ljubljana, Ljubljana, Slovenia}


\begin{abstract}
We calculate a nonequilibrium steady state of a quantum XX chain in the presence of dephasing and driving due to baths at chain ends. The obtained state is exact in the limit of weak driving while the expressions for one- and two-point correlations are exact for an arbitrary driving strength. In the steady state the magnetization profile and the spin current display diffusive behavior. Spin-spin correlation function on the other hand has long-range correlations which though decay to zero in either the thermodynamical limit or for equilibrium driving. At zero dephasing a nonequilibrium phase transition occurs from a ballistic transport having short-range correlations to a diffusive transport with long-range correlations. 
\end{abstract}



\maketitle
{\em Introduction--}
The theory of equilibrium processes is well developed with known equilibrium distribution functions. Nonequilibrium physics is on the other hand much more difficult to treat. One of the simpler nonequilibrium situations is the steady state setting to which the system relaxes after a long time under nonequilibrium driving. Still, it is not even known if a generic distribution function for a nonequilibrium steady state (NESS) exists. Particulary important aspect of NESS is the question of transport, for instance, it is not fully understood under what conditions does one have a diffusive transport~\cite{Lebowitz}. There is also a large disparity between classical and quantum systems. For many classical models, like various exclusion processes~\cite{exclusion}, an explicit solution is known, greatly facilitating our understanding of interesting phenomena that can occur in NESS. Physics of quantum NESS is on the other hand still a rather uncharted territory with only a few exact solutions. One such class of exactly solvable systems are those described by master equations quadratic in fermionic operators. NESS in doubly-infinite XY chain has been studied in~\cite{Araki}. Quadratic open systems can in fact be exactly diagonalized in operator space~\cite{prosen:njp} making it possible to study nonequilibrium phase transitions~\cite{iztok:08}. XX chain interacting with baths has been analytically studied also in~\cite{Karevski:09}. Time evolution of quadratic open systems can as well be calculated efficiently in terms of matrix product operators~\cite{clark:10}, similarly as for closed systems~\cite{pre:07}. Tight-binding (ie., XX) model in the presence of the environment has been studied in~\cite{esposito:05}. Superoperator corresponding to the master equation has been diagonalized and found that there is a crossover from nondiffusive to diffusive behavior as the chain size is increased. Diffusion in an XX model with classical noise has been considered in~\cite{amir:09}. In a recent work~\cite{temme:09} quantum exclusion processes, described by the master equation with nearest-neighbor Lindblad operators for stochastic jumps as well as with coherent unitary XX part, have been numerically studied. For some parameters similar long-range correlations are obtained in the NESS as here.

In the present work we are going to provide an explicit solution for the NESS of a quantum model, showing that the system exhibits diffusive transport and has long-range correlations. Furthermore, as a parameter of the model is varied a nonequilibrium phase transition occurs. To our knowledge this is the first quantum model for which one is able to analytically show diffusive behavior.

The system studied is a one-dimensional XX chain of spin-(1/2) particles,
\begin{equation}
H=\sum_{j=1}^{n-1} (\sigma_j^{\rm x} \sigma_{j+1}^{\rm x} +\sigma_j^{\rm y} \sigma_{j+1}^{\rm y}).
\label{eq:H}  
\end{equation}
To induce a nonequilibrium situation we couple the system at both ends to baths at different potentials. In addition, each spin is also exposed to a dephasing. The same model has been numerically studied in~\cite{njp}. The evolution of system's density matrix is governed by the master equation of the Lindblad form~\cite{lindblad},
\begin{equation}
\frac{{\rm d}}{{\rm d}t}{\rho}=\ii [ \rho,H ]+ {\cal L}^{\rm bath}(\rho)+{\cal L}^{\rm deph}(\rho)={\cal L}(\rho).
\label{eq:Lin}
\end{equation}
Nonunitary terms ${\cal L}^{\rm bath}$ and ${\cal L}^{\rm deph}$ are expressed in terms of Lindblad operators as $\sum_k \left( [ L_k \rho,L_k^\dagger ]+[ L_k,\rho L_k^{\dagger} ] \right)$. The bath superoperator ${\cal L}^{\rm bath}={\cal L}^{\rm bath}_{\rm L}+{\cal L}^{\rm bath}_{\rm R}$ is a sum of Lindblad terms for the left and the right most spin. Each consists of two Lindblad operators $L^{\rm L/R}_{1,2}$,
\begin{eqnarray}
L^{\rm L}_1=\sqrt{\Gamma} \frac{\sqrt{1-\mu}}{2}\sigma^+_1,\quad L^{\rm L}_2&=&\sqrt{\Gamma} \frac{\sqrt{1+\mu}}{2} \sigma^-_1, \nonumber \\
L^{\rm R}_1 = \sqrt{\Gamma} \frac{\sqrt{1+\mu}}{2}\sigma^+_n,\quad L^{\rm R}_2&=&\sqrt{\Gamma} \frac{\sqrt{1-\mu}}{2} \sigma^-_n,
\label{eq:Lbath}
\end{eqnarray}
$\sigma^\pm=\sigma^{\rm x} \pm {\rm i}\, \sigma^{\rm y}$. Two parameters for baths are the coupling strength $\Gamma$ and the driving strength $\mu$. Dephasing acts independently on each spin separately, ${\cal L}^{\rm deph}=\sum_{j=1}^n{{\cal L}^{\rm deph}_j}$, with ${\cal L}^{\rm deph}_j$ having only one Lindblad operator $L^{\rm deph}_j=\sqrt{\frac{\gamma}{2}}\sigma^{\rm z}_j$. It is diagonal in the Pauli basis, ${\cal L}^{\rm deph}_j(\sigma^{\rm x/y}_j)=-2\gamma\, \sigma^{\rm x/y}_j$, while ${\cal L}^{\rm deph}_j(\sigma^{\rm z}_j)=0$ and ${\cal L}^{\rm deph}_j(\mathbbm{1}_j)=0$. Dephasing with strength $\gamma$ therefore causes an exponential decay of the off-diagonal elements in the diagonal basis of $\sigma^{\rm z}$.

To find the NESS of the master equation (\ref{eq:Lin}) we are going to write $\rho$ as a sum over $4^n$ different products of Pauli matrices and solve for unknown coefficients by demanding stationarity ${\cal L}(\rho)=0$. Using Jordan-Wigner transformation we can express the whole system in terms of spinless fermions. While the $H$ (\ref{eq:H}) and ${\cal L}^{\rm bath}$ (\ref{eq:Lbath}) are quadratic in fermionic operators, ${\cal L}^{\rm deph}$ is not. Because it involves a product of two $\sigma^{\rm z}$ it is quartic and therefore the recently introduced~\cite{prosen:njp,iztok:08} analytical solution for open quadratic systems can not be used. Note that all quadratic systems are ballistic.

Let us start with an unnormalized ansatz,
\begin{equation}
\rho \sim \mathbbm{1}+\mu( A + B) +\frac{\mu^2}{2}\left(A B +B A \right)+\mu^2(C +D + F)+{\cal O}(\mu^3),
\label{eq:ansatz}
\end{equation}
with the individual terms being
\begin{equation}
\mu A= \sum_{j=1}^n a_j \sigma^{\rm z}_j,\qquad \mu B=\frac{b}{2}\sum_{k=1}^{n-1}j_k,
\end{equation}
where $j_k=2(\sigma_k^{\rm x} \sigma_{k+1}^{\rm y}-\sigma_k^{\rm y} \sigma_{k+1}^{\rm x})$ is the operator of local spin current, while the other two terms are
\begin{eqnarray}
\mu^2 C&=&\sum_{j=1}^n \sum_{k=j+1}^n (C_{j,k}+a_j a_k) \sigma_j^{\rm z} \sigma_k^{\rm z},\\
\mu^2 D&=&\sum_{j=1}^{n-2} \frac{d_j}{2} \left(\sum_{l=j+1}^{n-1}\sigma^{\rm z}_j j_l-\sum_{l=1}^{n-1-j}j_l \sigma^{\rm z}_{n+1-j} \right),\\
\mu^2 F&=&\frac{f}{8}\sum_{k,l=1 \atop k\neq l}^{n-1} j_k j_l.
\end{eqnarray}
We shall show that using an appropriate coefficients in the above ansatz, acting with ${\cal L}$ on such $\rho$ gives zero up-to ${\cal O}(\mu^3)$, ie., that the above ansatz is the correct solution up-to 3rd order in the driving strength $\mu$. Powers of $\mu$ written in front of various coefficients in the ansatz suggest a scaling of the corresponding coefficient in the solution, that is, post festum we shall see that indeed $\mu A \propto \mu$, $\mu^2 C \propto \mu^2$, etc.. Note that for equilibrium driving, $\mu=0$, the solution is $\rho \sim \mathbbm{1}$, which can be interpreted as an infinite temperature state, although one must be careful speaking about the temperature because integrable systems in general do not thermalize, even when coupled to reservoirs~\cite{thermalization}.

Demanding that the expansion coefficients in front of all operators in ${\cal L}(\rho)$ vanish one obtains a set of equations. Coefficients in front of two boundary operators $\sigma^{\rm z}_{1,n}$ give
\begin{equation}
-b-\Gamma \mu-\Gamma a_1=0,\qquad b+\Gamma\mu-\Gamma a_n=0.
\label{eq:z1zn}
\end{equation}
These two equations are exact to all orders in $\mu$ because one can get $\sigma_{1,n}^{\rm z}$ term in ${\cal L}(\rho)$ only from terms of the form $\mathbbm{1},\sigma_{1,n}^{\rm z}$ or $j_{1,n}$ in $\rho$, which are all included in our ansatz (\ref{eq:ansatz})! That is, $\tr{[\sigma_{1,n}^{\rm z}{\cal L}(\rho) ]}\neq 0$ only if $\rho \in \{\mathbbm{1},\sigma_{1,n}^{\rm z},j_{1,n} \}$. As we shall see, all equations we are going to use are, in fact, exact to all orders in $\mu$! Equations (\ref{eq:z1zn}) give relation $a_n=-a_1$ and $a_1=-\mu-\frac{b}{\Gamma}$. From coefficients in front of $\sigma_{j}^{\rm z}$ in the bulk ($j\neq 1,n$) one gets that the expectation of the local current $j_k$ is independent of the site $k$, ie. $b_k=b$, which has already been explicitly taken into account in our ansatz.

Equations which we get by demanding that the coefficients in front of $j_i$ are zero are as follows:
\begin{equation}
(a_i-a_{i+1})-(h^{(3)}_i-h^{(3)}_{i-1})-b \Upsilon_i^{(2)}=0,
\label{eq:jk}
\end{equation}
with $\Upsilon_i^{(k)}=2\gamma+\Gamma \delta_{i,1}+\Gamma \delta_{i+k-1,n}$, while $h^{(3)}_j$ is the coefficient in the expansion of $\rho$ in front of the operator $H^{(3)}_j=\sigma_j^{\rm x} \sigma_{j+1}^{\rm z} \sigma_{j+2}^{\rm x}+\sigma_j^{\rm y} \sigma_{j+1}^{\rm z}\sigma_{j+2}^{\rm y}$. At the boundaries, terms from operators having site indices out of chain range $1 \le j \le n$ are absent. These equations (\ref{eq:jk}) are exact because we can get $j_k$ term only from $j_k,\sigma_k^{\rm z}$ or $\sigma_{k-1}^{\rm x}\sigma_{k}^{\rm z}\sigma_{k+1}^{\rm x}+\sigma_{k-1}^{\rm y}\sigma_{k}^{\rm z}\sigma_{k+1}^{\rm y}$. Before solving Eq.(\ref{eq:jk}) we are going to show that $h^{(3)}_i$ are actually zero.

To this end let us in addition to the expansion coefficients which are explicitly included in the ansatz, also denote by $h^{(k)}_j$ the expansion coefficient in front of the operator $H^{(k)}_j=\sigma_j^{\rm x} \sigma_{j+1}^{\rm z} \cdots \sigma_{j+k-2}^{\rm z} \sigma_{j+k-1}^{\rm x}+\sigma_j^{\rm y} \sigma_{j+1}^{\rm z} \cdots \sigma_{j+k-2}^{\rm z} \sigma_{j+k-1}^{\rm y}$ and by $b^{(k)}_j$ the expansion coefficient in front of $B^{(k)}_j=\sigma_j^{\rm x} \sigma_{j+1}^{\rm z} \cdots \sigma_{j+k-2}^{\rm z} \sigma_{j+k-1}^{\rm y}-\sigma_j^{\rm y} \sigma_{j+1}^{\rm z} \cdots \sigma_{j+k-2}^{\rm z} \sigma_{j+k-1}^{\rm x}$. $B^{(k)}_j$ is a generalization of the spin current while $H^{(k)}_j$ is the hopping operator to the $(k-1)$th neighbor. Note that the rotational symmetry of ${\cal L}$ around $z$-axis imposes the value of the two signs in both operators. For $b^{(2)}_k$ we already know that it is site-independent and equal to $b$, in addition we define $b^{(1)}_k \equiv 0$. We are going to show that all $h^{(k)}_j$ as well as $b^{(k)}_j$ (apart from $b^{(2)}_j$) are zero. Demanding that the coefficients in front of $H^{(k)}_j$ in ${\cal L}(\rho)$ are zero, we get for $k=2,\ldots,n$ the equations
\begin{equation}
(b^{(k+1)}_j-b^{(k+1)}_{j-1})+(b^{(k-1)}_j-b^{(k-1)}_{j+1})-h^{(k)}_j \Upsilon_j^{(k)}=0.
\label{eq:hk}
\end{equation}
Equations from the coefficients in front of $B^{(k)}_j$ are, on the other hand, for $k=3,\ldots,n$ 
\begin{equation}
(h^{(k+1)}_j-h^{(k+1)}_{j-1})+(h^{(k-1)}_j-h^{(k-1)}_{j+1})+b^{(k)}_j \Upsilon_j^{(k)}=0.
\label{eq:bk}
\end{equation}
With exact equations (\ref{eq:hk}) and (\ref{eq:bk}) we have a closed set of exactly as many equations as there are unknown $h^{(k)}_j$ and $b^{(k)}_j$. Homogeneous set is nonsingular with the only solution being trivial $h^{(k)}_j=0$ and $b^{(k)}_j=0$. This means that NESS $\rho$ does not contain any $B^{(k)}_j$ terms with $k>2$ nor any $H^{(k)}_j$ terms with $k \ge 2$.

Going now back to the equations obtained from currents (\ref{eq:jk}), they can be solved exactly. Together with two equations (\ref{eq:z1zn}) they give the solution for $b$ and all $a_j$,
\begin{equation}
b=-\frac{\mu}{\Gamma+\frac{1}{\Gamma}+(n-1)\gamma},
\label{eq:ab0}
\end{equation}
\begin{eqnarray}
a_1&=&-\frac{b}{\Gamma}-\mu \nonumber \\
a_2&=&-b(\frac{1}{\Gamma}+\Gamma+2\gamma)-\mu \nonumber \\
a_3&=&-b(\frac{1}{\Gamma}+\Gamma+4\gamma)-\mu \nonumber \\
&\vdots & \nonumber\\
a_{n-1}&=&-b(\frac{1}{\Gamma}+\Gamma+2(n-2)\gamma)-\mu\nonumber\\
a_{n}&=&-b(\frac{1}{\Gamma}+2\Gamma+2(n-1)\gamma)-\mu.
\label{eq:ab}
\end{eqnarray}
This solution is exact to all orders in $\mu$ for any $n$, irrespective of other terms present in the expansion of $\rho$ not included in our ansatz. Before proceeding to solve for $C,D$ and $F$ let us for a moment discuss its physical significance. First, observe that the current and the magnetization expectation values in the NESS are simply equal to the corresponding coefficients, $\ave{\sigma_j^{\rm z}}=a_j$, $\ave{j}=2b$. Magnetization has a linear profile and the current scales as $j \sim \mu/n$ for large $n$ as long as $\gamma \neq 0$. Such behavior is typical for a system with normal (diffusive) transport. Spin conductivity is $\kappa = 1/\gamma$ and as one decreases the dephasing strength $\gamma$ the current diverges as $\sim 1/\gamma$~\cite{njp}. Also interesting, the limit of weak coupling to the bath, $\Gamma \to 0$, is at fixed $n$ and zero dephasing singular. That is, the current is proportional to $\Gamma$ as $j \sim \mu \Gamma$ while the magnetization at the boundaries is proportional to $\mu \Gamma^2$. In particular, without dephasing for $\gamma=0$, transport is ballistic with $b=-\Gamma \mu/(1+\Gamma^2)$, $a_1=-\Gamma^2 \mu/(1+\Gamma^2), a_n=\Gamma^2 \mu/(1+\Gamma^2)$, while $a_{j\neq 1,n}=0$, the same as in the recent result in~\cite{Karevski:09}.

To avoid cumbersome expressions for the 2nd order terms we from now on set $\gamma=\Gamma=1$. This has no qualitatively relevant consequences. In this case the solution (\ref{eq:ab},\ref{eq:ab0}) can be simply written as,
\begin{equation}
b=-\frac{\mu}{n+1},\quad a_j=-\frac{n+1-k_j}{n+1}\mu,\quad k_j=2j-\delta_{j,1}+\delta_{j,n}.
\label{eq:aib}
\end{equation}
Expansion coefficients in front of $\sigma_1^{\rm z} \sigma_k^{\rm z}$ give (using (\ref{eq:aib})),
\begin{eqnarray}
3b^2+d_1+C_{1,2}&=&0,\qquad \hbox{for }k=2 \nonumber \\
d_{n-2}-C_{1,3}&=&0,\qquad \hbox{for }k=3 \nonumber\\
d_{n-3}-C_{1,4}&=&0,\qquad \hbox{for }k=4 \nonumber\\
&\vdots &\nonumber\\
d_{1}-C_{1,n}&=&0,\qquad \hbox{for }k=n.
\label{eq:z1zk}
\end{eqnarray}
Interior nearest-neighbor terms $\sigma_j^{\rm z} \sigma_{j+1}^{\rm z}$ give (non nearest-neighbor are zero)
\begin{equation}
2b^2+d_j+d_{n-j}=0,\qquad \hbox{for } j=1,\ldots,n/2.
\label{eq:zz}
\end{equation}
Equations (\ref{eq:z1zk},\ref{eq:zz}) are again exact because $\sigma_j^{\rm z} \sigma_{k}^{\rm z}$ term can only be obtained from terms of the same form or from $j_k \sigma_j^{\rm z}$ or $\sigma_j^{\rm z}$. Coefficients in front of $\sigma_1^{\rm z} j_k$, on the other hand, give
\begin{eqnarray}
f-2b^2+4d_1-C_{1,2}+C_{1,3}&=&0,\quad \hbox{for }k=2, \nonumber \\
2f-2b^2+4d_1-C_{1,k}+C_{1,k+1}&=&0,\quad 2<k<n-1, \nonumber \\
2f-2b^2+5d_1-C_{1,n-1}+C_{1,n}&=&0,\quad k=n-1.
\label{eq:z1jk}
\end{eqnarray}
Equations are again exact for the same reason as before. The set of equations (\ref{eq:z1zk},\ref{eq:zz},\ref{eq:z1jk}) can be solved, giving explicit expressions for $d_i$, $f$ and $C_{1,k}$. Using these in equations obtained for $\sigma_2^{\rm z}j_k$ enables one to obtained $C_{2,j}$, then from $\sigma_3^{\rm z}j_k$ we obtain $C_{3,j}$ and so on. At the end the solution is,
\begin{eqnarray}
f&=& \frac{b^2(n+1)}{n}=\frac{\mu^2}{n(n+1)} \nonumber \\
d_j&=&-\frac{b^2}{n} k_j=-\frac{\mu^2}{(n+1)^2n}k_j \nonumber \\
C_{i,j}&=& -\frac{b^2}{n} (k_i k_{n+1-j}+(n+1)\cdot\delta_{j,i+1}),
\label{eq:cdf}
\end{eqnarray}
where $k_j$ is the same is in $a_j$ (\ref{eq:aib}). The solution is exact to all orders in $\mu$. One can check that the ansatz with coefficients (\ref{eq:cdf}) and (\ref{eq:aib}) results in the ${\cal L}(\rho)$ having only terms of order ${\cal O}(\mu^3)$. One could systematically calculate also 3rd and higher order terms, however, because main physical picture already emerges from the 2nd order term we are going to stop here (due to algebraic structure all nonzero operators in ${\cal L}(\rho)$ are in fact products of three terms, each being either $\sigma_j^{\rm z}$ or $j_k$; more, 3rd order term in the ansatz would itself be a sum of such products of three operators, each being either $j_k$ or $\sigma_j^{\rm z}$).

Because coefficients in the ansatz are exact to all orders in $\mu$ the corresponding single- or two-point expectation values are also exact. Two-point connected correlation function of the magnetization, $C(i,j)=\ave{\sigma_i^{\rm z} \sigma_j^{\rm z}}-\ave{\sigma_i^{\rm z}}\ave{\sigma_j^{\rm z}}$, is symmetric across the diagonal and is for $j>i$ negative and equal to $C_{i,j}$ (\ref{eq:cdf}). On the diagonal we have $C(i,i)=1-a_i^2$. If we are away from the diagonal, $j\neq i+1$ and $i,j\neq 1,n$, the correlation function is, in terms of scaled variables, simply equal to $C(x=i/(n+1),y=j/(n+1))=-\frac{(2\mu)^2}{n} x(1-y)$.

\begin{figure}
  \centerline{\includegraphics[width=0.5\textwidth]{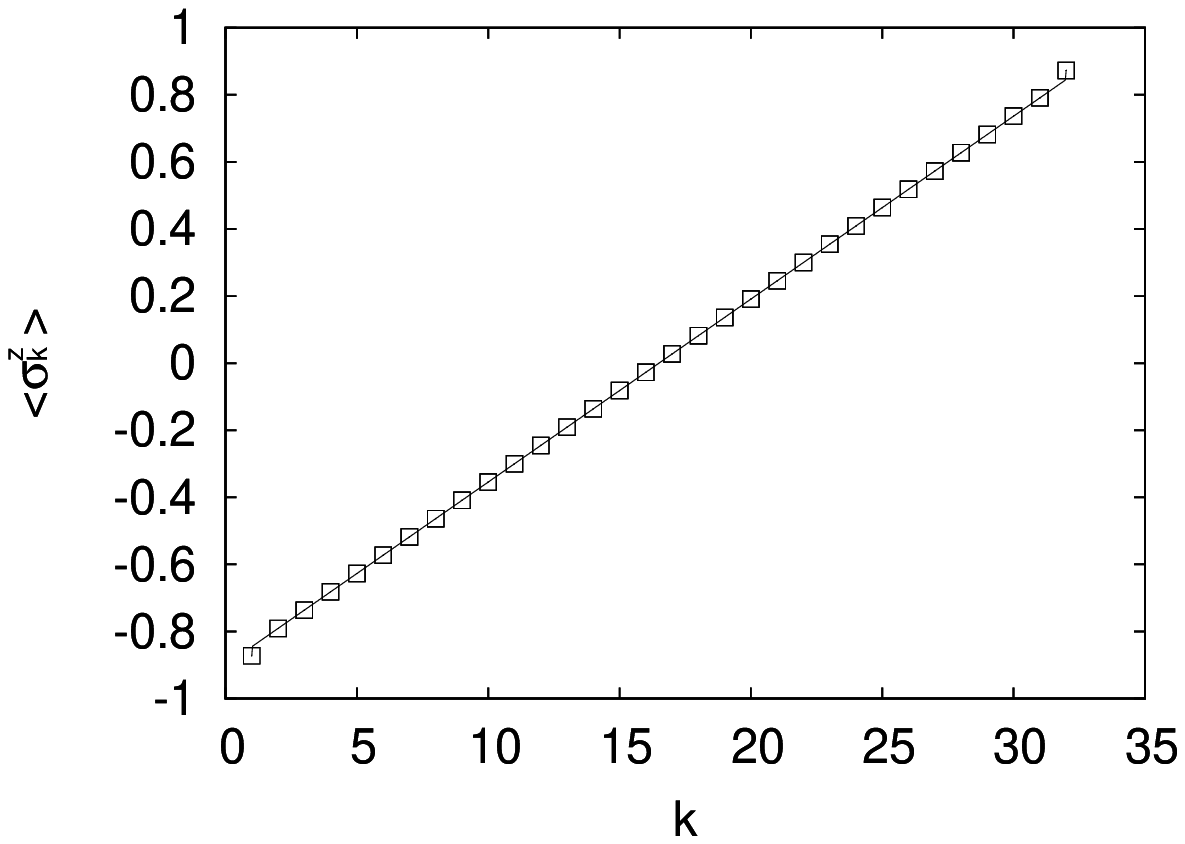}}
  \centerline{\includegraphics[width=0.5\textwidth]{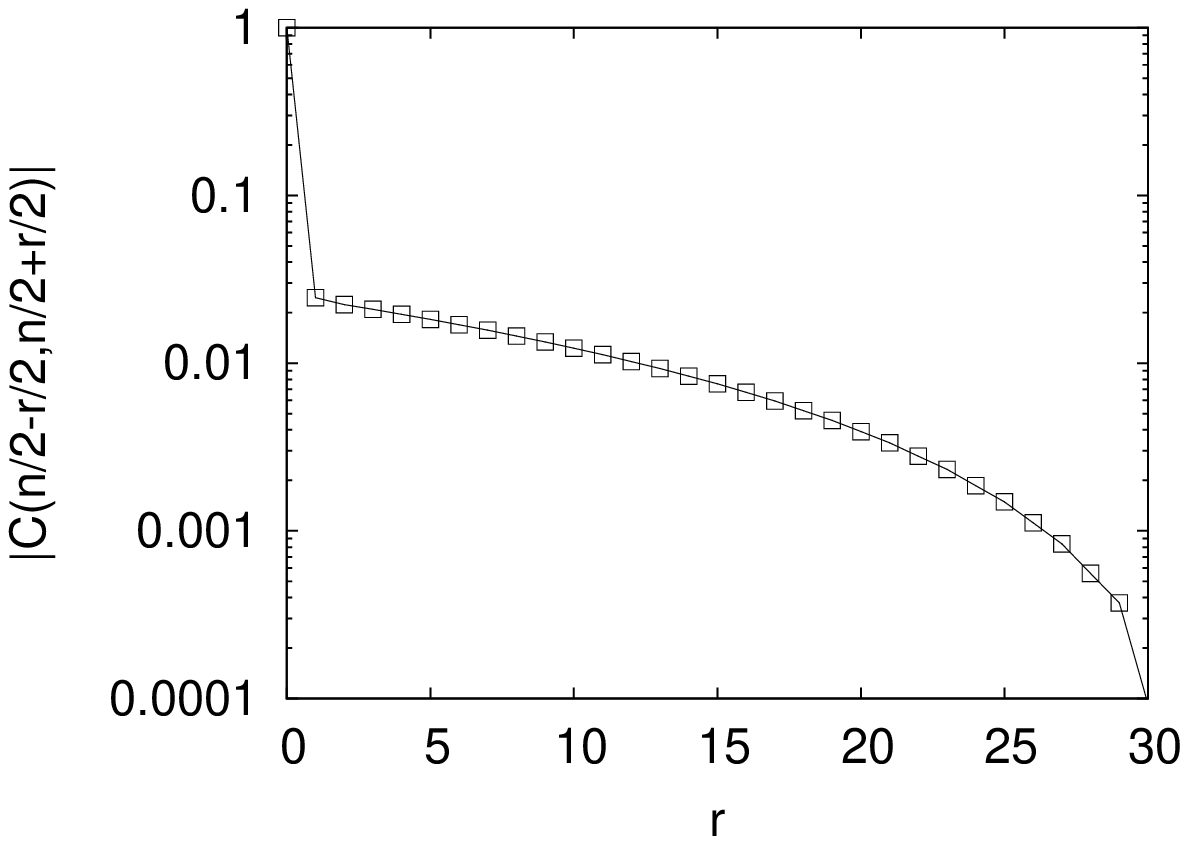}}
  \caption{Comparison between the numerical calculation (symbols) for $n=32$ and strong driving $\mu=0.9$ and the analytical solution (Eq.\ref{eq:aib},~\ref{eq:cdf}, full line). Top frame shows the magnetization profile and the bottom the correlation function.}
  \label{fig:z32}
\end{figure}

To independently verify our exact analytical solution for magnetization profile and correlation function we have compared it to the numerical simulation. Using the tDMRG~\footnote{For details of our implementation for master equations see~\cite{njp} or 3rd ref. under~\cite{heisenberg}.} we can calculate numerically exact expectation values in the NESS for chains with $n \sim 100$ spins. In Fig.~\ref{fig:z32} one can indeed see that the numerical and analytical solution agree perfectly also for very strong driving $\mu=0.9$.

Correlation function indicates that there is long-range order present, which though goes to zero either in the thermodynamic limit $n \to \infty$, or in the equilibrium limit $\mu \to 0$. Long-range order is therefore a finite size nonequilibrium phenomenon. Such correlations in the same XX model with dephasing have been numerically observed in~\cite{njp}. There, a qualitatively same behavior has been obtained also for an anisotropic XXZ model with dephasing, perhaps suggesting that the form of the solution found here is more general and would approximately apply also for other systems. In particular, we have indications~\cite{tobe} that a similar behavior is obtained for pure XXZ model without dephasing, whose seemingly normal spin transport~\cite{heisenberg} in the gapped regime is surprising and needs further understanding. It appears that long-range correlations are a rather common feature of quantum NESS~\cite{iztok:08,njp,temme:09,tobe}. The form of the correlation function $C(x,y)=-\frac{(2\mu)^2}{n} x(1-y)$ is the same as in many classical exclusion processes~\cite{exclusion}. Whether there exists a deeper connection with classical statistical models remains to be seen~\cite{temme:09}. All these results suggest that the form of the exact solution found here might have a more general validity. 

Let us for the end briefly discuss two-point correlations for the XX model without dephasing, $\gamma=0$. Calculation along the same lines as above results in the solution (for $\Gamma=1$) $f=\frac{\mu^2}{4}$, $d_j \equiv 0$, $C_{i,j}=-\frac{\mu^2}{4}\cdot\delta_{j,i+1}$, together with the already obtained $b=-\frac{\mu}{2}$ and $a_{j\neq1,n}=0$, while $a_1=-a_n=-\frac{\mu}{2}$. We can see that there are no long-range correlations present. Therefore, at $\gamma=0$ a nonequilibrium phase transition occurs from a ballistic transport with short-ranged correlations for $\gamma=0$ to a diffusive regime with long-range correlations for $\gamma \neq 0$.  

{\em Conclusion--} 
We have analytically calculated a nonequilibrium steady state of an open quantum spin chain to the 2nd order in the driving strength. One and two-point correlation functions are calculated exactly to all orders. The system, XX model with dephasing, shows diffusive transport with linear magnetization profile and current scaling as $\sim 1/n$. There are also long-range correlations present in the steady state. To our knowledge this is the first quantum system for which one is able to analytically show diffusive transport. At zero dephasing a nonequilibrium phase transition occurs from a short-ranged ballistic to a diffusive transport with long-range correlations. Furthermore, long-range correlations in the present model have the same form as in some classical statistical exclusion processes and there are some indications that the same qualitative behavior is found also in other quantum systems. Therefore, while the exact solution might be possible only for this special model, the results might have more general (linear response) validity. It is hoped that the presented result will lead to better understanding of quantum transport in one-dimensional systems as well as of the conditions under which long-range correlations appear in nonequilibrium quantum states. This could bridge a gap between relatively well explored classical exclusion precesses, where many analytical solutions are known, and quantum nonequilibrium steady states where almost no exact solutions are known. 
M\v Z is supported by the Program P1-0044 and the Grant J1-2208 of the Slovenian Research Agency.


\begin{thebibliography}{1}

\bibitem{Lebowitz} F.~Bonetto\etal{, J.~L.~Lebowitz, and L.~Rey-Bellet}, \tit{Fourier's law: A challenge to theorists} in {\em Mathematica Physics 2000}, A.~Fokas, A.~Grigoryan, T.~Kimble, and B.~Zegarlinski, eds. (Imperial College Press, London, 2000); {\tt arXiv:math-ph/0002052}

\bibitem{exclusion} R.~Stinchcombe, \tit{Stochastic non-equilibrium systems} 2001 Adv.~in Phys. {\bf 50} 431; B.~Derrida, \tit{Non-equilibrium steady states: fluctuations and large deviations of the density and of the current} 2007 J.~Stat.~Mech. P07023

\bibitem{Araki} H.~Araki and T.~G.~Ho, \tit{Asymptotic time evolution of a partitioned infinite two-sided isotropic XY-chain} 2000 Proc.~Steklov Inst.~Math. {\bf 228} 191; Y.~Ogata, \tit{Diffusion of magnetization profile in the XX model} 2002 Phys.~Rev.~E {\bf 66} 066123; W.~H.~Aschbacher and C.-A.~Pillet, \tit{Non-equilibrium steady states of the XY chain} 2003 J.~Stat.~Phys. {\bf 112} 1153; W.~Aschbacher and J.-M.~Barbaroux, \tit{Out of equilibrium correlations in the XY chain} 2006 Lett.~Math.~Phys. {\bf 77} 11

\bibitem{prosen:njp} T.~Prosen, \tit{Third quantization: a general method to solve master equations for quadratic open Fermi systems} 2008 New J.~Phys {\bf 10} 043026

\bibitem{iztok:08} T.~Prosen and I.~Pi\v zorn, \tit{Quantum phase transition in a far from equilibrium steady state of XY spin chain} 2008 Phys.~Rev.~Lett. {\bf 101} 105701; T.~Prosen and B.~\v Zunkovi\v c, \tit{Exact solution of Markovian master equations for quadratic fermi systems: thermal baths, open XY spin chains, and non-equilibrium phase transition} 2010 New J.~Phys {\bf 12} 025016

\bibitem{Karevski:09} D.~Karevski and T.~Platini, \tit{Quantum nonequilibrium steady states induced by repeated interactions} 2009 Phys.~Rev.~Lett. {\bf 102} 207207

\bibitem{clark:10} S.~R.~Clark\etal{, J.~Prior, M.~J.~Hartmann, D.~Jaksch, and
M.~B.~Plenio}, \tit{Exact matrix product solutions in the Heisenberg
picture of an open quantum spin chain} 2010 New J.~Phys {\bf 12} 025005

\bibitem{pre:07} T.~Prosen and M.~\v Znidari\v c, \tit{Is the efficiency of classical simulations of quantum dynamics related to integrability?} 2007 Phys.~Rev.~E {\bf 75} 015202(R); T.~Prosen and I.~Pi\v zorn, \tit{Operator space entanglement entropy in transverse Ising chain} 2007 Phys.~Rev.~A {\bf 76} 032316; I.~Pi\v zorn and T.~Prosen, \tit{Operator Space Entanglement Entropy in XY Spin Chains} 2009 Phys.~Rev.~B {\bf 79} 184416

\bibitem{esposito:05} M.~Esposito and P.~Gaspard, \tit{Emergence of diffusion in finite quantum systems} 2005 Phys.~Rev.~B {\bf 71} 214302

\bibitem{amir:09} A.~Amir\etal{, Y.~Lahini, and H.~B.~Perets}, \tit{Classical diffusion of a quantum particle in a noisy environment} 2009 Phys.~Rev.~E {\bf 79} 050105(R)

\bibitem{temme:09} K.~Temme\etal{, M.~M.~Wolf, and F.~Verstraete}, \tit{Stochastic exclusion processes versus coherent transport} 2009 {\tt arXiv:0912.0858}


\bibitem{thermalization} M.~\v Znidari\v c\etal{, T.~Prosen, G.~Benenti, G.~Casati, and D.~Rossini}, \tit{Thermalization and ergodicity in many-body open quantum systems} 2009 {\tt arXiv:0910.1075}

\bibitem{njp} M.~\v Znidari\v c, \tit{Dephasing-induced diffusive transport in the anisotropic Heisenberg model} 2010 New J.~Phys. {\bf 12} 043001

\bibitem{lindblad} G.~Lindblad, \tit{On the generators of quantum dynamical semigroups} 1976 Comm. Math. Phys. {\bf 48} 119; V.~Gorini\etal{, A.~Kossakowski, and E.~C.~G. Sudarshan}, \tit{Completely positive dynamical semigroups of N-level systems} 1976 J.~Math.~Phys. {\bf 17} 821; H.-P.~Breuer and F.~Petruccione, {\em The Theory of Open Quantum Systems} (Oxford University Press, Oxford, 2002)

\bibitem{tobe} T.~Prosen and M.~\v Znidari\v c, {\em in preparation}.

\bibitem{heisenberg} F.~Heidrich-Meisner\etal{, A.~Honecker, D.~C.~Cabra, and W.~Brenig}, \tit{Zero-frequency transport properties of one-dimensional spin-1/2 systems} 2003 Phys.~Rev.~B {\bf 68} 134436; M.~Michel\etal{, O.~Hess, H.~Wichterich, and J.~Gemmer}, \tit{Transport in open spin chains: A Monte Carlo wave-function approach} 2008 Phys.~Rev.~B {\bf 77} 104303; T.~Prosen and M.~\v Znidari\v c, \tit{Matrix product simulations of non-equilibrium steady states of quantum spin chains} 2009 J.~Stat.~Mech. P02035; S.~Langer\etal{, F.~Heidrich-Meisner, J.~Gemmer, I.~P.~McCulloch, and U.~Schollw\" ock}, \tit{Real-time study of diffusive and ballistic transport in spin-1/2 chains using the adaptive time-dependent density matrix renormalization group method} 2009 Phys.~Rev.~B {\bf 79} 214409; R.~Steinigeweg and J.~Gemmer, \tit{Density dynamics in translationally invariant spin-1/2 chains at high temperatures: A current-autocorrelation approach to finite time and length scales} 2009 Phys.~Rev.~B {\bf 80} 184402; J.~Sirker\etal{, R.~G.~Pereira, and I.~Affleck}\tit{, Diffusion and ballistic transport in one-dimensional quantum systems} 2009 Phys.~Rev.~Lett. {\bf 103} 216602

\end{thebibliography}
\end{document}